%% file: main.tex
\documentclass[twocolumn,showpacs,preprintnumbers,amsmath,amsfonts,amssymb,floatfix,aps,pra,superscriptaddress, bibnotes]{revtex4}

\usepackage{amssymb}
\usepackage{amsmath}
\usepackage{amsfonts}
\usepackage{bm}
\usepackage{color}
\usepackage{hyperref}
\usepackage{soul}
\usepackage[com pat=1.1.0]{tikz-feynman}
\usepackage{standalone}
\usetikzlibrary{arrows.meta,decorations.markings}
\usetikzlibrary{bending} 
\tikzset{
    arc arrow/.style args={%
    to pos #1 with length #2}{
    decoration={
        markings,
         mark=at position 0 with {\pgfextra{%
         \pgfmathsetmacro{\tmpArrowTime}{#2/(\pgfdecoratedpathlength)}
         \xdef\tmpArrowTime{\tmpArrowTime}}},
        mark=at position {#1-\tmpArrowTime} with {\coordinate(@1);},
        mark=at position {#1-2*\tmpArrowTime/3} with {\coordinate(@2);},
        mark=at position {#1-\tmpArrowTime/3} with {\coordinate(@3);},
        mark=at position {#1} with {\coordinate(@4);
        \draw[-{Latex[length=#2,bend]}]       
        (@1) .. controls (@2) and (@3) .. (@4);},
        },
     postaction=decorate,
     }
}

\newcommand{\be}{\begin{equation}}
\newcommand{\ee}{\end{equation}}

\usepackage{graphicx}

\begin{document}

\title{Exploring beyond-mean-field logarithmic divergences in Fermi-polaron energy}

\author{R. Alhyder}%
\email[Corresponding author: ]{ragheed.alhyder@ist.ac.at}
\affiliation{Center for Complex Quantum Systems, Department of Physics and Astronomy, Aarhus University, Ny Munkegade 120, DK-8000 Aarhus C, Denmark
}%
\affiliation{Institute of Science and Technology Austria (ISTA),
 Am Campus 1, 3400 Klosterneuburg, Austria
}%

\author{F. Chevy}%
\affiliation{Laboratoire de physique de l’Ecole Normale supérieure, ENS, Université PSL, CNRS, Sorbonne Université, Université de Paris, F-75005 Paris, France
}%
\affiliation{Institut Universitaire de France (IUF)
}%
\author{X. Leyronas}%
\affiliation{Laboratoire de physique de l’Ecole Normale supérieure, ENS, Université PSL, CNRS, Sorbonne Université, Université de Paris, F-75005 Paris, France
}%

\begin{abstract}
  We perform a diagrammatic analysis of the energy of a mobile impurity immersed in a strongly interacting two component Fermi gas to second order in the impurity-bath interaction. These corrections demonstrate divergent behavior in the limit of large impurity momentum. We show the fundamental processes responsible for these logarithmically divergent terms. 
  We study the problem in the general case without any assumptions regarding the fermion-fermion interactions in the bath. We show that the divergent term can be summed up to all orders in the Fermi-Fermi interaction and that the resulting expression is equivalent to the one obtained in the few body calculation.
  Finally, we provide a perturbative calculation to the second order in the Fermi-Fermi interaction in the annex, and we show the diagrams responsible for these terms.
\end{abstract}
\maketitle
\section{Introduction}
The physics of an impurity in a many-body ensemble is an intriguing problem, and constitutes a rich field of research in condensed matter physics. The study of the quantum impurity problem was initiated by Landau and Pekar who proposed that the properties of conduction electrons in a dielectric medium could be understood in terms of so-called polarons, i.e. quasi-particles resulting from the dressing of the electrons by a cloud of optical phonons of the surrounding crystal \cite{landau1948effective}.

More recently, the realization of spin and atomic mixtures of ultracold atoms have paved the way to the study of impurity problems in ultracold gases \cite{massignan2014polarons,Chevy2010Unitary}. In these systems, the impurity can be immersed in either a bosonic or a fermionic medium which leads to strikingly different behaviors and phenomena. On the one hand, since the low-lying excitation modes of a Bose-Einstein condensate are phonons, the Bose-polaron (an impurity immersed in a Bose-Einstein condensate) is quite similar to Landau-Pekar's polaron \cite{jorgensen2016}. By contrast, in the Fermi polaron case (an impurity immersed in a spin polarized gas of fermions \cite{Chevy2006,schirotzek2009ofp}) the impurity is dressed by a cloud of particle hole-pairs. 

The case of an impurity immersed in a spin $1/2$ superfluid was brought into the limelight following experimental works on Bose-Fermi superfluids \cite{Ferrier2014Mixture, roy2017two, yao2016observation}.  As the fermion-fermion interaction is varied in the BCS-BEC crossover, the fermionic background medium evolves from a weakly attractive interaction condensate of loosely bound Cooper pairs on the BCS (Bardeen-Cooper-Schrieffer) side of the crossover to a strongly attractive interaction on the BEC side where the Fermi gas condenses in a BEC of tightly-bound dimers. Thus, the polaronic state smoothly turn from a Fermi polaron on the BCS side to a Bose polaron on the BEC side of the crossover.

In the case of a zero-range coupling between the impurity and the fermions, a peculiar UV-divergent term appears when calculating the polaron energy perturbatively with respect to the impurity-fermion interaction \cite{Yi2015,Pierce19few}. This divergence is typical in three-body problems with contact interactions and was revealed first in the study of beyond mean-field corrections in dilute Bose-Einstein condensates \cite{wu1959ground}. Indeed, three-body bound states, the Efimov trimer states, have been studied in the case of the Bose polaron \cite{levinsen2015, Sun2017, grusdt2017}. This work is accomplished in the regime of the Born approximation with respect to the impurity-bath interaction, and therefore far from the universal regime where Efimov physics can form. However, the divergence has its origins in three-body physics and can be remedied using an effective field theory approach \cite{braaten1999quantum,braaten2002dilute}.
\st{It originates in three-body physics and can be remedied using an effective field theory approach} \cite{braaten1999quantum,braaten2002dilute}. 

In this scheme, the divergences are suppressed by introducing counter terms corresponding to effective three-body interactions \cite{Pierce19few}. However, this renormalization process only works if the density-density response of the fermionic superfluid obeys a specific scaling that was found to be incompatible with a mean-field description of the fermionic background. This inconsistency is due to the omission of the collective mode sector in the description of the excitation spectrum of the system in the mean-field approach \cite{Pierce19few,Bigue2022}. As a consequence, a proper regularization could be only be carried out within the framework of Random Phase Approximation (RPA) \cite{Bigue2022}.  

In this work, we compute  these divergent terms rigorously in the case of an imbalanced spin $1/2$ Fermi gas using Feynman diagrams from the density-density response function and we prove its behavior and relation to Tan's contact. Furthermore, we underpin the processes responsible for these divergent terms in the many-body problem without any assumptions regarding the fermion-fermion interactions in the bath. This is an important step in identifying diagrams which have signatures of the few-body physics in the problem and a step forward in understanding the polaron energy in the many-body problem.

We start by introducing the Hamiltonian of the system and a summary of the problem. Then we lay out the methodology used in identifying the dominant terms in the problem. We show that the processes responsible for the divergent behavior can be categorized in three families of diagrams which we provide concrete arguments for their expressions in the large impurity momentum limit.
\section{System description}
We consider the case of  an impurity immersed in a partially spin-polarized double Fermi sea at zero temperature. The imbalance between the two  spin populations is chosen beyond the threshold for Clogston-Chandrasekhar transition \cite{clogston1962ulc,chandrasekhar1962,Chevy2010Unitary}, allowing us to disregard diagrams where anamolous propagators play a role. Furthermore, we assume that the impurity-fermion interaction is weak and attractive, thus we can treat it perturbatively.

Introducing a quantization volume $\mathcal V$, the Hamiltonian of the system is written as:
\begin{align}
\begin{split}
	\hat{H} 
	&= \sum_{\textbf{k},\sigma} \varepsilon_k \hat{a}^\dagger_{\textbf{k},\sigma} \hat{a}_{\textbf{k},\sigma}  
	+ \sum_{\textbf{q}}\varepsilon^{(i)}_q\hat{c}^\dagger_{\textbf{q}} \hat{c}_{\textbf{q}}
	\\&+ \frac{g'_0}{\mathcal{V}}\sum_{\textbf{k},\textbf{q},\textbf{k}',\textbf{q}',\sigma}\delta_{\textbf{k}+\textbf{q},\textbf{k}'+\textbf{q}'}\hat{c}^\dagger_{\textbf{q'}}\hat{a}^\dagger_{\textbf{k'},\sigma} \hat{c}_{\textbf{q}} \hat{a}_{\textbf{k},\sigma}
	\\& +\frac{g_0}{\mathcal{V}}\sum_{\textbf{k},\textbf{q},\textbf{k}',\textbf{q}'} \delta_{\textbf{k}+\textbf{q},\textbf{k}'+\textbf{q}'}\hat{a}^\dagger_{\textbf{k'},\uparrow}\hat{a}^\dagger_{\textbf{q'},\downarrow} \hat{a}_{\textbf{q},\downarrow} \hat{a}_{\textbf{k},\uparrow},
	\label{eq:HamMat}
\end{split}
\end{align}
where $\hat{a}_{\textbf{k},\sigma}$ is the annihilation operator of a fermion with momentum $\textbf k$ and spin $\sigma$, $\hat{c}_{\textbf{q}}$ is the annihilation operator of an impurity with momentum $\textbf q$. Noting $m_i$ and $m$ the respective masses of the impurity and of the fermions \st{respectively},  $ \varepsilon_k=(\hbar^2 k^2)/(2m)$ is the kinetic energy of a fermion with wavevector ${\bf k}$ and $\varepsilon^{(i)}_q=(\hbar^2 q^2)/(2m_i)$ is the kinetic of the impurity with wavevector ${\bf q}$. $g'_0$ and $g_0$ are the bare coupling constants of the fermion-impurity and the fermion-fermion interactions respectively.
\\
The coupling constant $g'_0$ is related to the scattering length and the cut-off $\Lambda$ through the following equation:
\begin{equation}
	\frac{1}{g'_0} = \frac{1}{g'} - \frac{1}{\mathcal{V}}\sum_{k<\Lambda} \frac{2m_r}{\hbar^2 k^2},
	\label{eq:CCMat}
\end{equation}
where $g'$ is the physical coupling constant between the impurity and background fermions. It is related to the scattering length $a$ using the relation: $g' = 2\pi \hbar^2 a'/m_r$, with $m_r=(m\,m_i)/(m+m_i)$ the impurity-fermion reduced mass. \\
By using perturbation theory we can obtain an expression for the polaron energy up to second order \cite{Pierce19few, Bigue2022}:
\begin{equation}
	E_{\text{pol}} =g' n + \frac{{g'}^2n }{\mathcal{V}}\sum_{\textbf{q}<\Lambda} [ \frac{2m_r}{\hbar^2 q^2}-\chi_1(\textbf{q},\epsilon^{(i)}_{\textbf{q}})],
	\label{eq:EnerPolMat}
\end{equation}
where
\be
\chi_1(\bm q,E)=\frac{1}{N}\sum_{\alpha}\frac{\left|\langle \alpha|\hat{n}_{{\bf q}}|0\rangle\right|^2}{(E+E_\alpha-E_0)}.
\label{eqdefchiRqE}
\ee
Here $\hat{n}_{{\bf q}}=\sum_{{\bf k},\sigma}a^{\dagger}_{{\bf k},\sigma}a^{\phantom{}}_{{\bf k}+{\bf q},\sigma}$. $|0\rangle$ is the ground state of the interacting bath and $\{|\alpha\rangle\}$ denotes a basis of eigenvectors the Hamiltonian of the fermionic bath alone.

It was conjectured in \cite{Pierce19few} that in the large momentum limit 
\be
\chi_1({\bf q},\varepsilon^{\rm (i)}_q)-\frac{1}{\varepsilon^{\rm (r)}_q}={\cal O}\left(\frac{1}{q^3}\right),
\ee
thus leading to a logarithmically divergent value of the sum appearing in Eq. (\ref{eqdefchiRqE}). This divergence can be healed by introducing a three-body interaction \cite{Pierce19few} but only under the assumption thet  $\chi$ obeys the following asymptotic behavior 
\be
\chi_1({\bf q},\varepsilon^{\rm (i)}_q)\underset{q\rightarrow\infty}{=}\frac{1}{\varepsilon^{\rm (r)}_q}\left[1-\pi^2\kappa(\eta)\frac{m}{m_r}\frac{\mathcal{C}_2}{Nq}+...\right],
\label{Eq:AsymptoticChi}
\ee
where $\mathcal{C}_2$ is Tan's contact parameter of the many-body background \cite{tan2008large}, $\eta=m_{\rm i}/m$  and 
\begin{align}
\begin{split}
	\kappa(\eta) &= \kappa_I(\eta) + \kappa_{II}(\eta) + \kappa_{III}(\eta) 
	\\&=  \frac{\sqrt{\eta^3(\eta + 2)}}{2 \pi^3 (\eta + 1 )^2}
	- \frac{\eta}{2 \pi^3} \arctan\left({\frac{1}{\sqrt{\eta(\eta + 2)}}}\right)
	\\& - \frac{4}{\pi^3} \sqrt{\frac{\eta}{\eta + 2}} \arctan\left({\sqrt{\frac{\eta}{\eta + 2}}}\right)^2,
	\label{eq:KappaEta}
\end{split}
\end{align}
The addition of a diverging term was initially done as a conjecture with the goal of regularizing the expression in the large impurity momentum limit.
This conjecture is supported  by an RPA analysis of the excitation modes of the fermionic background \cite{Bigue2022} and the purpose of the present article is to prove rigorously this behavior by solving the full many-body problem and studying the processes responsible for the divergence. After a first introduction of the methodology in Part \ref{sec:methodology}, 
 we will recover Eqs. (\ref{Eq:AsymptoticChi}) and (\ref{eq:KappaEta}) in Part \ref{sec:generalcase} using scaling arguments to generalize the results to an arbitrary order of the interaction parameter. This is supported by a perturbative analysis in the bare fermion-fermion coupling constant to identify the elementary processes by calculating exactly all diagrams contributions to second order in Appendix \ref{AppendixA}.
\vspace{-0.55cm}
\section{Methodology}
\label{sec:methodology}
\begin{figure}[h]
\centering
\includegraphics[width=0.35\textwidth]{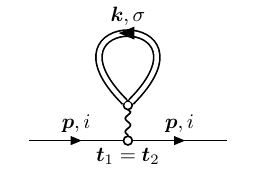}
\caption{The lowest order diagram for the impurity self-energy in momentum-time representation. The empty circle represents the bare impurity-fermion coupling constant $g'_0$. The double line represents the exact fermion Green's function $G_\sigma$. The single lines represents the impurity's Green's function $G_i$.
}
\label{Fig-Sigmai_1}
\end{figure}

\begin{figure}[h]
\centering
\includegraphics[width=0.36\textwidth]{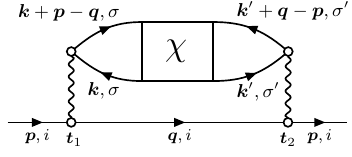}
\caption{The diagram for the impurity self-energy in momentum-time representation at second order in $g'_0$. The rectangle represents the exact density-density response function of the fermionic many-body background. 
}
\label{Fig-Sigmai_2}
\end{figure}

The chemical potential of the impurity or equivalently the binding energy $E_{pol}$ of the polaron is given by the self-energy of the impurity at zero momentum \cite{Combescot2007}
\begin{equation}
E_{pol}=\Sigma_i({\bf 0},E_{pol}).
\label{eq:epolsigma}
\end{equation}
We compute $\Sigma_i$ perturbatively in $\hat{H}_{int}$ up to second order, {\it i.e.} up to order $(g'_0)^2$. Note that we do not treat perturbatively the fermion-fermion interaction. The two diagrams up to order $(g'_0)^2$ are shown in Figs.(\ref{Fig-Sigmai_1}) and (\ref{Fig-Sigmai_2}). 
In time and momentum representation, the impurity self-energy at first order reads
\begin{equation}
\Sigma_i^{(1)}({\bf p},t_2-t_1)=\frac{g'_0}{\mathcal{V}}\sum_{{\bf k},\sigma}(-i)
G_{\sigma}({\bf k},0^-)\delta(t_2-t_1),
 \end{equation}
 where $G_{\sigma}({\bf k},t)$ is the exact fermion Green's function. Using $\sum_{{\bf k},\sigma}(-i)
G_{\sigma}({\bf k},0^-)/\mathcal{V}=n$, the total density, we find, after taking the time Fourier transform
\begin{equation}
\Sigma^{(1)}_i({\bf p},E)=g'_0\,n.\label{eq:Sigmai_1}
\end{equation}

The self-energy at second order is written in a diagrammatic form in Fig. (\ref{Fig-Sigmai_2}), therefore it can be written as follows
\begin{eqnarray}
\Sigma_i^{(2)}({\bf p},t_2-t_1)&=&\left(\frac{g'_0}{\mathcal{V}}\right)^2
\sum_{{\bf q}}
(-i)e^{-i\varepsilon^{(i)}_{ q}(t_2-t_1)}\times\nonumber\\
&&\langle
\hat{n}_{{\bf p}-{\bf q}}(t_2)
\hat{n}_{{\bf q}-{\bf p}}(t_1)
\rangle\Theta(t_2-t_1)\label{eq:sigma2t},
\end{eqnarray}
where we have used that the free Green's function of the impurity is $G^{(0)}_i({\bf q},t)=(-i)\Theta(t)\exp(-i\varepsilon^{(i)}_q t)$. Taking the time Fourier transform of Eq.(\ref{eq:sigma2t}), we obtain
\begin{align}
  \begin{split}
\Sigma_i^{(2)}({\bf p},E)
&
=
\left(\frac{g'_0}{\mathcal{V}}\right)^2
\sum_{{\bf q},\alpha}
\frac{
|\langle\alpha|\hat{n}_{{\bf q-p}} |0\rangle|^2}
{
E-\varepsilon^{(i)}_q-E_{\alpha}+E_0+i\,0^+
}\\
&=-\frac{{g_0'}^2n }{\mathcal{V}}\sum_{q<\Lambda} \chi_1(\textbf{q},\epsilon^{(i)}_{\textbf{q}}),
\label{eq:Sigmai_2_omega}
\end{split}
\end{align}
At order $(g'_0)^2$, we must solve, using Eq.(\ref{eq:epolsigma}), $E_{pol}=\Sigma_i^{(1)}({\bf 0},E_{pol})+\Sigma_i^{(2)}({\bf 0},E_{pol})$. 
At lowest order, we can replace $E_{pol}$ by $0$ in $\Sigma_i^{(2)}({\bf 0},E_{pol})$. We express $g'_0$ in terms of $g'$ by expanding Eq.(\ref{eq:CCMat}): 
$$g'_0=g'+\frac{{g'}^2}{\mathcal{V}}\left(\sum'_{{\bf q}}\frac{2\,m^*}{q^2}+\cdots\right).$$ 
In this way, we find that $\Sigma_i^{(1)}$ gives the first term in Eq.(\ref{eq:EnerPolMat}), and the first term in the sum. At same order, $g'_0$ can be simply replaced by $g'$ in $\Sigma_i^{(2)}$ and this term provides the contribution associated with the function $\chi_1$ in the sum appearing in Eq. (\ref{eq:EnerPolMat}). Our analysis shows that with the two diagrams of Figs.(\ref{Fig-Sigmai_1}, \ref{Fig-Sigmai_2}), we recover Eq.(\ref{eq:EnerPolMat}).

The response function $\chi_1$ is directly related to the time-ordered density-density response function $\chi$ through
\begin{equation}
\chi_1({\bf q},\varepsilon^{(i)}_q)=\frac{-1}{N}\int_0^{+\infty}e^{-i \varepsilon^{(i)}_q t}\chi(-\textbf{q},t) dt,
\label{eq:chi1chiT}
\end{equation}
where
\begin{equation}
\chi(\textbf{q},t) = -i \langle T[n_{\textbf{q}}(t) n_{-\textbf{q}}] \rangle.
\end{equation}
here, $T$ is the time ordering operator. 

The goal is to prove the asymptotic behavior Eq.(\ref{Eq:AsymptoticChi}), and in order to do so, we will classify the Feynman diagrams contributing to the diagram in $\chi(\textbf{q},t)$, hence, $\lim_{|\bm{q}| \rightarrow \infty}\chi_1({\bf q},\varepsilon^{(i)}_q)$.

These results hold for the case of an imbalanced Fermi gas at zero temperature. We work on expanding the diagram in Fig. \ref{Fig-Sigmai_2} using the bold diagrams formalism \cite{prokofev2008bdm, vanHoucke2012feynman, rossi2018}, in order to find the contributions that scale as $1/q^3$ in the $q\rightarrow \infty$ limit, i.e. the ones responsible for remedying the divergent term found in $\chi_1(\bm{q},E)$. 
\\This is achieved by taking into account the cases where: a) no interaction vertices are involved, b) one fermionic interaction vertex is present, c) two fermionic interaction vertices are present in the fermionic bubble. We take inspiration for this procedure from generalizing the perturbative diagrams we have studied extensively, see Appendix \ref{AppendixA}. In general, as is explained in Appendix \ref{appendixC}, we do not expect any other diagrams to contribute to the diverging term. Furthermore, since no restrictions are made on the interaction nature between the fermionic particles in the bath, these results hold beyond the BCS regime.
\section{Computing the dominant diagrams}
\label{sec:generalcase}
In the weakly interacting limit between fermions, we study in Appendix \ref{AppendixA} the perturbative diagrams contributions to the divergent terms and we show the calculation for one of these contributions fully.

Here, we focus on the more general case, where no assumptions are made about the fermion-fermion interaction strength, we make qualitative arguments to distinguish the contributions to the diverging term.
We can classify the diagrams contributing $\chi_1({\bf q},\varepsilon^{(i)}_q)$ in three families for the diagrams of $\chi({\bf q},t>0)$ using the bold diagrams formalism \cite{prokofev2008bdm, vanHoucke2012feynman, rossi2018}. The bold diagrams formalism is a diagrammatic approach to the many-body problem where non-interacting propagators are replaced by the fully dressed Green's functions $G$ lines resulting from the resummation series, and the two-body interaction vertices are replaced by the fully dressed two-body vertices $\Gamma$. 
The bold diagrams are then classified in three families depending on the number of bold vertices $\Gamma$ they contain. The first family contains diagrams with no bold vertices, the second family contains diagrams with one bold vertex and the third family contains diagrams with two bold vertices. We detail the contributions of each family in the following.

We define a "typical" energy scale $E_{t y p}=k_{t y p}^{2} /(2 m)$
with the wave vector amplitude $k_{t y p}=\max (\left|a^{-1}\right|,|m \mu|^{1 / 2})$. 
In the $q\gg k_{t y p}$ limit, we also define a cut-off $\epsilon$ in time: $\epsilon \ll 1/E_{t y p}$ and $q^{2} / m \,\epsilon \gg 1$ and a cut-off $\Lambda$ in momentum: $\Lambda \gg k_{t y p}$ and $\Lambda \ll q$.

We will use the following properties of exact Green's function $G_{\sigma}({\bf k},t)$ and exact two-particle vertices 
$\Gamma({\bf P},t)$ of interacting fermions of the bath  (see \cite{vanhoucke2019diagmc} for these properties in imaginary time):

\textit{Property 1:} If $|p| \gg k_{typ}, 
G_{\sigma}(\mathbf{p}, t)$ is small, except in a small interval $0 \leq t \lesssim(2 m) / p^{2}$, where it tends to the Green's function of a particle in vacuum: $G_{\sigma}(\mathbf{p}, t)\simeq-i\Theta(t) e^{-i \frac{p^{2}}{2 m} t}$.
\\
\textit{Property 2:}  If $0<t \ll t_{t y p}, G_{\sigma}(\mathbf{p},-t) \rightarrow i n_{\mathbf{p},\sigma}$, where $n_{\mathbf{p},\sigma}=\left\langle c_{\mathbf{p}, \sigma}^{\dagger} c_{\mathbf{p}, \sigma}\right\rangle$ is the occupation number of the mode $\mathbf{p}, \sigma$.
\\
\textit{Property 3:} if $k \gg k_{t y p}$ and 
$t\lesssim m/k^2$
, we can write to leading order $ G_{\sigma}(\mathbf{k},-t)\simeq i \frac{\mathcal{C}_{2}}{k^{4}} e^{-i \frac{k^{2}}{2 m} t}$
where $\mathcal{C}_2$ is Tan's contact per unit volume.\\
\textit{Property 4:} If $|\mathbf{P}| \gg k_{typ}
, \Gamma(\mathbf{P}, t)$ is small, except in a small time interval $0 \leq t \lesssim(4 m) / P^{2}$ where it tends to the vertex of two particles in vacuum: $\Gamma(\mathbf{P}, t) \simeq \Gamma_{v a c}(\mathbf{P}, t)=-4\sqrt{\frac{\pi}{m^3 t}}e^{i\frac{\pi}{4}} e^{-i \frac{P^{2}}{4 m} t} \Theta(t)$.
\\
\textit{Property 5:} If $0 \leq t \ll t_{t y p}$, $\int \frac{d^{3} P}{(2 \pi)^{3}} \Gamma(\mathbf{P},-t) \simeq -i\, \mathcal{C}_{2} / m^{2}$. 

\subsection{ No interaction vertices }

The bold diagram for $\chi(\mathbf{q}, t)$ with no two-particle vertex $\Gamma$ is simply a bubble diagram with the exact Green's functions.
\begin{figure}[h]
\centering
\includegraphics[width=0.16\textwidth]{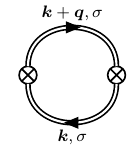}
\caption{The bold diagram with no vertex $\Gamma$ that contributes to the density-density response function.
}
\label{gen1}
\end{figure} 
In momentum and time variables, this diagram is given by (Fig. \ref{gen1}):
\begin{equation}
  {\chi}^T_{a}\left(\textbf{q},t\right)=-i
  \sum_{{\bf k},\sigma}
  G_{\sigma}(\mathbf{k}+\mathbf{q}, t) G_{\sigma}(\mathbf{k},-t)
\end{equation}
From Eq.(\ref{eq:chi1chiT}), we find the corresponding contribution to $\chi_1$:
\begin{align}
\begin{split}
\chi_{1\,a}\left(\textbf{q},\varepsilon^{(i)}_q\right)&=\sum_{\sigma}\frac{i}{n}\int_{0}^{+\infty}\hspace{-.5 cm}dt 
  \int \frac{d^{3} k}{(2 \pi)^{3}} 
  e^{-i \varepsilon^{(i)}_q t} G_{\sigma}(\mathbf{k}-\mathbf{q}, t) \\&G_{\sigma}(\mathbf{k},-t).
  \label{eq:chi1a}
  \end{split}
\end{align}
$G_{\sigma}(\mathbf{k}, t)$ is the Green's function of a fermion with momentum $\mathbf{k}$ and spin $\sigma$. We have used $\sum_{{\bf k}}\to V\,\int d^{3} k/(2 \pi)^{3}$ and $n=N/V$ the total density ($V$ is the volume).


First, in Eq. \eqref{eq:chi1a} consider the contribution $t<\epsilon$ and $k<\Lambda$ in the integrals. 
Since $q \gg \Lambda \geq k$, we can replace $G_{\sigma}(\mathbf{k}+\mathbf{q}, t)$ 
by $G_{\sigma}(\mathbf{q}, t)$ at lowest order. 
If we use the first property from above, we can write $G_{\sigma}(\mathbf{q}, t) \simeq -i\Theta(t) e^{-i \frac{q^{2}}{2 m} t}$ at lowest order. 
Since $t<\epsilon \ll t_{t y p}$, we can use the second property and replace at lowest order $G_{\sigma}(\mathbf{k},-t)$ by $i n_{\mathbf{k}}$. Here, the integral on $ \int_{|\mathbf{k}|<\Lambda} \frac{d^{3} k}{(2 \pi)^{3}} n_{\mathbf{k},\sigma}$ tends to $n_{\sigma}$ in the limit $\Lambda / k_{t y p} \rightarrow \infty$. We can perform the time integral on $\int_{0}^{\epsilon} d t e^{-i \frac{q^{2}}{2 m_{r}} t}=(-i)\left(1-e^{-i \frac{q^{2}}{2 m_{r}} \epsilon}\right)\left(2 m_{r}\right) / q^{2}$. The phase $\frac{q^{2}}{2 m_{r}} \epsilon \gg 1$ gives a fast oscillating term that we can neglect.

As a conclusion, the small time, small wave vector contribution to Eq. \eqref{eq:chi1a} gives $\frac{2 m_{r}}{q^{2}}$, in the $|\textbf{q}| \rightarrow \infty$ limit .
\\
Second, we subtract the term of order $q^{-2}$ in Eq. \eqref{eq:chi1a} 
\begin{align}
  \begin{split}
&\chi_{1\,a}\left(\textbf{q},\varepsilon^{(i)}_q\right)-\frac{2 m_{r}}{q^{2}} \\=&
\frac{1}{n}\sum_{\sigma}\int_{0}^{+\infty} d t\, e^{-i \frac{q^{2}}{2 m_{i}}t}  \frac{1}{i} \int \frac{d^{3} k}{(2 \pi)^{3}}\Big(G_{\sigma}(\mathbf{k}-\mathbf{q}, t) G_{\sigma}(\mathbf{k},-t)\\&-\frac{1}{i} e^{-i \frac{q^{2}}{2 m} t} G_{\sigma}\left(\mathbf{k}, 0^{-}\right)\Big)
\label{eq:chi1a2}
\raisetag{10pt}
\end{split}
\end{align}
In the integral on the RHS of Eq.\eqref{eq:chi1a2}, we evaluate the contribution of the domain $\{t \in[0, \epsilon],|\mathbf{k}|>\Lambda\}$. In this domain, we can use \textit{property 1} and replace $G_{\sigma}(\mathbf{k}+\mathbf{q}, t)$ with $-i e^{-i \frac{(\mathbf{k}+\mathbf{q})^{2}}{2 m} t}$. We use  \textit{property 3} and replace $G_{\sigma}(\mathbf{k},-t)$ with $i \frac{\mathcal{C}_{2}}{k^{4}} e^{-i \frac{k^{2}}{2 m} t}$ and $G_{\sigma}\left(\mathbf{k}, 0^{-}\right)$with $i \frac{\mathcal{C}_{2}}{k^{4}}$. We find
\begin{align*}
  \begin{split}
-i \int_{|\mathbf{k}|>\Lambda} \frac{d^{3} k}{(2 \pi)^{3}} \frac{\mathcal{C}_{2}}{k^{4}} \int_{0}^{\epsilon} d t\left[e^{-i\left(\frac{q^{2}}{2 m_{i}}+\frac{(\mathbf{k}+\mathbf{q})^{2}}{2 m}+\frac{k^{2}}{2 m}\right) t}-e^{-i \frac{q^{2}}{2 m_{r}} t}\right].
\end{split}
\end{align*}
Neglecting fast oscillating terms, we find for the time integral : $(-i)\left(\frac{1}{\frac{q^{2}}{2 m_{i}}+\frac{(\mathbf{k}+\mathbf{q})^{2}}{2 m}+\frac{k^{2}}{2 m}}-\frac{2 m_{r}}{q^{2}}\right)$.

Finally, the wave vector integral can be performed after the change of variable $\mathbf{k} \rightarrow q \mathbf{k}$. The lower bound for the norm of $\mathbf{k}$ is $\Lambda / q$, that we set to zero at lowest order. The volume element scales like $q^{3}$ and the integrand like $q^{-6}$. This gives the $q^{-3}$ dependence. As a conclusion, in the $|\textbf{q}| \rightarrow \infty$ limit, the integral on the rhs of Eq. \eqref{eq:chi1a2}  gives the contribution
$$
-\frac{\left(4 m \,\mathcal{C}_{2}\, J_{a}(\eta)\right)}{n}\frac{1}{q^{3}} 
$$
where $(\hat{q}$ is a unit vector $)$
\begin{align*}
  \begin{split}
    J_{a}(\eta)&=\int \frac{d^{3} k}{(2 \pi)^{3}} \frac{1}{k^{4}}\left(\frac{\eta}{\eta+1}-\frac{1}{\frac{1}{\eta}+(\mathbf{k}+\hat{q})^{2}+k^{2}}\right)\\&=\frac{1}{4 \pi} \frac{\sqrt{\eta^{3}(\eta+2)}}{(\eta+1)^{2}}
  \end{split}
\end{align*}
This is the dominant contribution in the $|\textbf{q}| \rightarrow \infty$ limit as is explained below and we find
$$
\chi_{1\, a}\left(q,\frac{q^{2}}{2 m_{i}}\right)=\frac{2 m_{r}}{q^{2}}\left( 1-\frac{m}{m_r}\frac{\mathcal{C}_{2}}{n}\frac{1}{2\pi} \frac{\sqrt{\eta^{3}(\eta+2)}}{(\eta+1)^{2}} \frac{1}{q}+\cdots\right)
$$
Since $\mathcal{C}_2/n=C_2/N$, we recover in the second term the $\kappa_I(\eta)$ contribution of Eqs. \eqref{Eq:AsymptoticChi} and \eqref{eq:KappaEta}.
\subsection{ One interaction vertex }
\begin{figure}[h]
\centering
\includegraphics[width=0.4\textwidth]{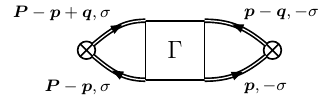}
\caption{The diagram with one interaction vertex and bold propagators for the fermions.
}
\label{gen2}
\end{figure}

The only diagram with one bold $\Gamma$ is the diagram show in Fig. \ref{gen2}.
The analytic expression for 
$\chi_{1\,b}(\mathbf{q}, \varepsilon^{(i)}_q)$ is 
 \begin{align}
 \begin{split}
\chi_{1\,b}(\mathbf{q}, \varepsilon^{(i)}_q)&=\frac{1}{n}\sum_{\sigma}\int_0^{+\infty}\hspace{-.5cm}dt\int \frac{d^{3} p}{(2 \pi)^{3}} \int \frac{d^{3} P}{(2 \pi)^{3}} \int_{-\infty}^{+\infty}\hspace{-.5cm} d t_{1} \int_{-\infty}^{+\infty}\hspace{-.5cm} d t_{2} \\&
 e^{-i \frac{q^{2}}{2 m_{i}}t}\,G_{\sigma}\left(\mathbf{P}-\mathbf{p}-\mathbf{q}, t_{1}\right) G_{\sigma}\left(\mathbf{P}-\mathbf{p},-t_{2}\right)\\& \Gamma\left(\mathbf{P}, t_{2}-t_{1}\right) G_{-\sigma}\left(\mathbf{p}, t-t_{2}\right) G_{-\sigma}\left(\mathbf{p}+\mathbf{q}, t_{1}-t\right)\\
 \raisetag{12pt}
 \label{eq:chi1b}
 \end{split}
 \end{align}
In case $|\mathbf{P}|>\Lambda$, from \textit{property 4} we see that the dominant contribution to $\Gamma$ is for $t_2 -t_1 >0$, which contradicts the time-ordering from the dominant contributions for Green's functions where $t_1>0$ and $t_2<0$. So we can neglect the contribution of the domain $|\mathbf{P}|>\Lambda$ in the integral on $\mathbf{P}$.\\
In case $|\mathbf{P}|<\Lambda$, and all fermionic wave vectors larger than $\Lambda$, due to \textit{property 1}, we can replace the Green's functions by their vacuum values. In the $|\textbf{q}| \rightarrow \infty$ limit, since the momenta are large compared to $|\mathbf{P}|$, we can set $\mathbf{P}=\mathbf{0}$ in the Green's functions. Due to the retarded nature of the fermionic Green's functions, we have the time ordering: $t_{1}>0,-t_{2}>0$, $t-t_{2}>0$ and $t_{1}-t>0$. For $t>0$, the integration domain is $\left\{\left(t_{1}, t_{2}\right), t_{1}>t, t_{2}<0\right\}$. Since all fermionic wave-vectors are large, the dominant contributions in the time integrals come from small time differences smaller than $\epsilon$. The time argument $t_{2}-t_{1}$ in $\Gamma$ is negative and much smaller than $t_{\text {typ }}$. At lowest order, we can replace $\Gamma\left(\mathbf{P}, t_{2}-t_{1}\right)$ by $\Gamma\left(\mathbf{P}, 0^{-}\right)$. We define time differences $t_{2}^{\prime}=-t_{2}$ and $t_{1}^{\prime}=t_{1}-t$ which vary between 0 and $\epsilon$. In the $|\textbf{q}| \rightarrow \infty$ limit, the time $t$ also lies between 0 and $\epsilon$, due to the $e^{-i \frac{q^{2}}{2 m_{i}} t}$ in the Fourier transform of $\chi_{b}^{T}(\mathbf{q}, t)$. We have the exponential term

$$
e^{-i\left(\frac{q^{2}}{2 m_{i}}+\frac{p^{2}}{2 m}+\frac{(\mathbf{p}+\mathbf{q})^{2}}{2 m}\right) t} e^{-i\left(\frac{(\mathbf{p}+\mathbf{q})^{2}}{m}\right) t_{1}^{\prime}} e^{-i\left(\frac{(\mathbf{p})^{2}}{m}\right) t_{2}^{\prime}}
$$
In Eq. \eqref{eq:chi1b}, the integrals on times give
$$
(-i)^{3} \frac{1}{\frac{q^{2}}{2 m_{i}}+\frac{p^{2}}{2 m}+\frac{(\mathbf{p}+\mathbf{q})^{2}}{2 m}} \frac{1}{\frac{(\mathbf{p}+\mathbf{q})^{2}}{m}} \frac{1}{\frac{(\mathbf{p})^{2}}{m}}
$$
Due to {\it property 5}, the integral on ${\bf P}$ gives a factor $-i\,\mathcal{C}_2/m^2$.
Finally, after rescaling of $\mathbf{p}$ by $q$, we also find a $q^{-3}$ scaling and the asymptotic behavior for $\chi_{1\,b}\left(\mathbf{q},\frac{q^{2}}{2 m_{i}}\right)$
$$
-\frac{\left(4 m \mathcal{C}_{2}\,J_{b}(\eta)\right)}{n}\frac{1}{q^{3}} ,
$$
where
\begin{align*}
\begin{split}
  J_{b}(\eta)&=-\int \frac{d^{3} p}{(2 \pi)^{3}} \frac{1}{\frac{1}{\eta}+p^{2}+(\mathbf{p}+\hat{q})^{2}} \frac{1}{(\mathbf{p}+\hat{q})^{2}} \frac{1}{p^{2}}\\&=-\frac{1}{4 \pi} \eta \arctan \left(\frac{1}{\sqrt{\eta(\eta+2)}}\right)
  \end{split}
\end{align*}
This is also the dominant contribution in the $|\textbf{q}| \rightarrow \infty$ limit and we obtain the result
$$
\chi_{1\,b}\left(\mathbf{q},\frac{q^{2}}{2 m_{i}}\right)=\frac{2m_r}{q^2}\,
\frac{m}{m_r}\frac{\mathcal{C}_{2}}{n}\frac{1}{q}
\frac{\eta}{2\pi} \arctan \left(\frac{1}{\sqrt{\eta(\eta+2)}}
\right)
+\cdots
$$
This is equal to the $\kappa_{II}(\eta)$ contribution in Eqs. \eqref{Eq:AsymptoticChi} and \eqref{eq:KappaEta}.
\subsection{Two interaction vertices}\label{sec2vert}
\begin{figure}[h]
\centering
\includegraphics[width=0.36\textwidth]{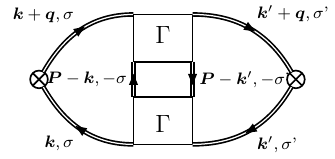}
\caption{The diagram with two bold interaction vertices and bold propagators for the fermions.
}
\label{gen3}
\end{figure}

The diagrams with two bold vertices have the form in Fig. \ref{gen3} with different permutations of the fermionic lines.
All permutations do not contribute except for the previous diagram and the one where the two fermions connecting the interaction vertices have different spins, which has the same exact contribution resulting in a factor $2$ in the final result. To this order, there exists  another possible diagram that we argue in Appendix \ref{appendixB} that it doesn't contribute to the divergent term. The analytic expression for this diagram is
\begin{align}
	\begin{split}
    \chi_{c}^{T}(\mathbf{q}, t)&=-i\sum_{\sigma,\sigma'}\int_{\mathbb{R}^{4}} \prod_{i=1}^{4} d t_{i} \int \frac{d^{3} k d^{3} k^{\prime} d^{3} P}{(2 \pi)^{9}} G_{\sigma}\left(\mathbf{k}+\mathbf{q}, t_{1}\right)
    \\& G_{\sigma}\left(\mathbf{k}, -t_{4}\right) G_{-\sigma}\left(\mathbf{P}-\mathbf{k}, t_{1}-t_{4}\right) \Gamma\left(\mathbf{P}+\mathbf{q}, t_{2}-t_{1}\right) \\&
  G_{\sigma'}\left(\mathbf{k}'+\mathbf{q}, t-t_{2}\right) G_{\sigma'}\left(\mathbf{k}', t_{3}-t\right)  
  \\& G_{-\sigma'}\left(\mathbf{P}-\mathbf{k}', t_{3}-t_{2}\right)\Gamma\left(\mathbf{P}, t_{4}-t_{3}\right)
  \label{eq:chiTc}
  \raisetag{10pt}
\end{split}
\end{align}
and $\chi_{1\,c}(\mathbf{q}, \varepsilon^{(i)}_q)$ as its time domain Fourier transform.
In the $|\textbf{q}| \rightarrow \infty$ limit, we assume the dominant contribution to the integral comes from the high wave-vectors regions. In these regions, due to \textit{property 1} from the previous calculation, the time arguments are restricted to small positive values and the Green's functions can be replaced by vacuum values. This implies that we have the time-ordering: $t_{1}>0>t_{4}$ and $t_{3}>t>t_{2}$. 
\\Next, we assume that one the two momenta of the interaction vertices are smaller than $\Lambda$. This means that we have two possibilities: either case $a)$, $|\mathbf{P}-\mathbf{q}|<\Lambda$ and $|\mathbf{P}| \rightarrow \infty$, or case $b)$ $|\mathbf{P}|<\Lambda$ and $|\mathbf{P}-\mathbf{q}| \rightarrow \infty$.
In case $a)$, due to \textit{property 4}, we have $t_{4}-t_{3}>0$. This is in contradiction with the time-ordering $t_{4}<0$ and $t_{3}>t>0$, and therefore we must exclude this case.\\
In case $b)$, since $|\mathbf{P}|$ is bounded and all fermionic wave vectors tend to infinity, we can set $\mathbf{P}=\mathbf{0}$ in all of the Green's functions at lowest order. This means that $|\mathbf{P}-\mathbf{q}| \rightarrow \infty$ and we can replace $\Gamma\left(\mathbf{P}-\mathbf{q}, t_{2}-t_{1}\right)$ by $\Gamma_{v a c}\left(\mathbf{P}-\mathbf{q}, t_{2}-t_{1}\right) \simeq \Gamma_{v a c}\left(-\mathbf{q}, t_{2}-t_{1}\right)$ due to \textit{property 4} and the fact that $|\mathbf{P}|$ is bounded.
\\
We define time differences which are all positive: $\tau_1=t_1$, $\tau_2=t_2-t_1$, $\tau_3=t_3-t$, $\tau_4=-t_4$ and $\tau_5=t-t_2$. These time differences must be of the order of the inverse of typical kinetic energies, which are of the order $m/q^2\ll t_{typ}$. As a consequence, we can replace the time difference $t_4-t_3$ by $0^-$ in $\Gamma({\bf P},t_4-t_3)$.
The integral on $\mathbf{P}$ gives $\int_{|\mathbf{P}|<\Lambda}\frac{d^{3} P}{(2 \pi)^{3}} \Gamma\left(\mathbf{P}, t_{4}-t_{3} \rightarrow 0^{-}\right)=\Gamma\left(\mathbf{r}=\mathbf{0}, t=0^{-}\right)=i \frac{\mathcal{C}_{2}}{m^{2}}$, where we have used the fact that $\Lambda \gg k_{t y p}$ and extend the wave-vector integral to all space. 
 Neglecting fast oscillating terms as before, the integrals on time differences $\left\{\tau_{i}\right\}$ gives 
\begin{align*}
\begin{split}
 &\frac{1}{\frac{k^{\prime 2}}{m}} 
  \frac{1}{\frac{\left(\mathbf{k}^{\prime}+\mathbf{q}\right)^{2}}{2 m}
  +\frac{\left(\mathbf{k}^{\prime}\right)^{2}}{2 m}+\frac{\mathbf{q}^{2}}{2 m_{i}}}
\int_{0}^{+\infty} d\tau_2 
e^{-i \frac{q^{2}}{2 m_{i}} \tau_2} \Gamma_{v a c}\left(-\mathbf{q}, \tau_2\right) \\
  &\frac{1}{\frac{k^{2}}{m}} 
  \frac{1}{\frac{(\mathbf{k}+\mathbf{q})^{2}}{2 m}+\frac{(\mathbf{k})^{2}}{2 m}+\frac{\mathbf{q}^{2}}{2 m_{i}}}
  \end{split}
\end{align*}
The upper bound on $\tau_2$ is $\epsilon$, but since $\epsilon q^{2} /\left(2 m_{i}\right) \gg 1$, we have extended it to infinity. The integral on $\tau_2$ can be performed analytically and is equal to $-\frac{8 \pi}{m} \sqrt{\frac{\eta}{\eta+2}} \frac{1}{q} \propto \frac{1}{q}$.

The integrals on $\mathbf{k}^{\prime}$ and $\mathbf{k}$ are performed after the change of variables $\mathbf{k}^{\prime} \rightarrow q \mathbf{k}^{\prime}$ and $\mathbf{k} \rightarrow q \mathbf{k}$. After this change of variables, we can set the lower bound $\Lambda / q$ to zero at lowest order. For each integral, a factor $q^{3}$ comes from the volume element and a factor $q^{-4}$ comes from the integrand, which makes the integral scales like $q^{-1}$. Together with the $q^{-1}$ scaling of the intermediate $\Gamma_{v a c}$, we recover the $q^{-3}$ dependence.

Putting together all the factors, we find for the $|\textbf{q}| \rightarrow \infty$ limit of the diagram in \ref{eq:chi1c} the contribution to $\chi_{1\,c}\left(q,\frac{q^{2}}{2 m_{i}}\right)$
$$
\frac{128\, \pi\, m\, \mathcal{C}_{2}}{n}\sqrt{\frac{\eta}{\eta+2}}\left(J_{c}(\eta)\right)^{2}
\frac{1}{q^3}$$
where
$$
J_{c}(\eta)=\int \frac{d^{3} k}{(2 \pi)^{3}} \frac{1}{k^{2}} \frac{1}{\frac{1}{\eta}+k^{2}+(\mathbf{k}+\hat{q})^{2}}=\frac{1}{4 \pi} \arctan \sqrt{\frac{\eta}{\eta+2}}
$$
We recover the $\kappa_{III}(\eta)$ contribution in Eqs. \eqref{Eq:AsymptoticChi} and \eqref{eq:KappaEta}.

In the Appendix \ref{appendixC}, we give arguments which justify that diagrams with more than 3 interaction vertices give subleading contributions to $\chi_{1}\left(q,\frac{q^{2}}{2 m_{i}}\right)$ in the $q\to\infty$ limit.

\section{Conclusion}
We have calculated the leading order contribution to the static density response function $\chi(\mathbf{q}, t)$ in the limit of large momentum transfer $|\mathbf{q}| \rightarrow \infty$ for an impurity immersed in a two-component Fermi gas with contact interaction. 
\\We have shown that the leading order contribution is given by the sum of three bold diagrams. The first diagram is a bubble diagram with the exact Green's functions. The second diagram contains one interaction vertex and the third diagram contains two interaction vertices. The leading order contribution to $\chi(\mathbf{q}, t)$ in the $|\mathbf{q}| \rightarrow \infty$ limit is given by the sum of the three bold diagrams. 
\\This helps shed light on the origin of such logarithmic divergences and provides a motivation to calculate these contribution in other cases such as the Bose polaron to see if the same behavior is present. This systematic approach has been done at zero temperature, but can also be performed at finite temperature using the same methods where similar results are expected.
\acknowledgments{
We thank Félix Werner and Kris Van Houcke for interesting discussions. 
}

\appendix
\input{SupplementalMaterial.tex}

\bibliographystyle{unsrt}
\bibliography{libraryUsed.bib}

\end{document}

%% file: SupplementalMaterial.tex
\section[short]{Perturbative limit}
\label{AppendixA}
\begin{figure}[h]
\centering
\includegraphics[width=0.32\textwidth]{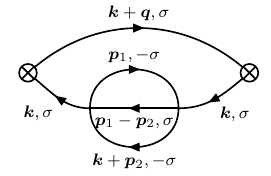}
\caption{A contribution to density-density response function to second order in the fermion-fermion interaction. This is the only diagram from the self-energy insertion diagrams which includes a divergent term that has a $1/q^3$ scaling.
}
\label{pert1}
\end{figure}

For this approach, we will compute the response function $\chi({\bf q},\omega)$ in wave vector and frequency. We will use the formula
\begin{equation}
\chi_{1}({\bf q},\varepsilon_q^{(i)})=\frac{i}{N}\int_{-\infty}^{+\infty}
\frac{d\omega}{2\pi}\frac{\chi(-{\bf q},\omega)}{\omega+\varepsilon_q^{(i)}-i\,0^+}
\end{equation}
In the perturbative limit, we consider the case where the fermion-fermion interaction is weak, i.e. $g_0 \rightarrow 0^{-}$. In this perturbative limit, Tan's contact per unit volume is given by $\mathcal{C}_2=m^2\,g_0^2\,n_{\downarrow}\,n_{\uparrow}$.
In the zeroth order, the only diagram that contributes to the polaron energy is a bubble diagram that scales as $1/q^2$ to the leading order with no $q^{-3}$ term. To first order, the contribution of the dumbell and tadpole diagrams in the large momentum limit are of $1/q^4$ order. This is expected since the diverging term is second order in the Fermi-Fermi interaction $g_0$.
To second order in $g_0$, the diagrams that contribute to the polaron energy consist of different families which we detail in the following:

\textbf{1.} Diagrams where the two resulting fermions from the interaction with the impurity in diagram \ref{Fig-Sigmai_2} don't interact with each other. These are called the self-energy insertion diagrams since in one of the fermionic lines, we introduce two first order diagrams or one second order diagram from the self energy of the impurity. By computing all these contributions, we find that the only contributing diagram is the one in Fig \ref{pert1}. In this Appendix, we show the expression for this diagram and we prove that it gives the contribution $\kappa_{I}(\eta)$ in Eq. \eqref{eq:KappaEta}.
The diagram in Fig. \ref{pert1} can be computed using the following equation:
\begin{align}
	\begin{split}
g_0^2&\sum_{k,p_1,p_2,\sigma} 
G_{0,\sigma}(k)^2
G_{0,-\sigma}(p_1)
G_{0,\sigma}(p_1-p_2)
\\&
G_{0,-\sigma}(k+p_2)
G_{0,\sigma}(k+q)
\end{split}
\end{align} 
Remembering that all diagrams should be advanced with respect to the frequency $\omega$ we write the Green's functions product in the following manner:
\begin{align}
	\begin{split}
&\sum_{k,p_1,p_2} 
\frac{1}{(\omega_1-E_{\bm{k}}+i\eta_1)^2}
\frac{1}{\nu_1-E_{\bm{p}_1}-i\eta}
\\& 
\frac{1}{\nu_1-\nu_2-E_{\bm{p}_1-\bm{p}_2}+i\eta}
\frac{1}{\omega_1+\nu_2-E_{\bm{k}+\bm{p}_2}+i\eta}
\\&
\frac{\theta(|\bm{k}+\bm{q}|>k_F)}{\omega_1+\omega-E_{\bm{k}+\bm{q}}+i0^+}
\label{eq:1.6}
	\end{split}
\end{align}
where the sign of $\eta$ and $\eta_1$ determines the boundaries for the amplitudes of the wave-vectors $\bm{p}_1$, $\bm{p}_2$ and $\bm{k}$. We can first perform the integration over the frequencies $\nu_1$ and $\nu_2$:
\begin{align}
	\begin{split}
&\sum_{k, \bm{p}_1, \bm{p}_2}
\frac{\theta(|\bm{k}+\bm{q}|>k_F)}{(\omega_1-E_{\bm{k}}+i\eta_1)^2(\omega_1+E_{\bm{p}_1}-E_{\bm{p}_1-\bm{p}_2}-E_{\bm{k}+\bm{p}_2}+i\eta)}
\\&\frac{1}{(\omega_1+\omega-E_{\bm{k}+\bm{q}}+i0^+)}
+\frac{1}{2\epsilon_{p_2}}
\label{eq:1.6}
	\end{split}
\end{align}
We found out two contributions that give the $1/q^3$ behavior:
\begin{align}
\begin{split}
&\eta_1=0^+ ,\, \eta = 0^- \Rightarrow |\bm{k}|>k_F, \, |\bm{p}_1|>k_F, \, |\bm{p}_1-\bm{p}_2|<k_F,\\& |\bm{k}+\bm{p}_2|<k_F
\raisetag{12pt}
\label{firstcase}
\end{split}
\end{align}
\begin{align}
\begin{split}
&\eta_1=0^- ,\, \eta = 0^+ \Rightarrow |\bm{k}|<k_F, \, |\bm{p}_1|<k_F, \, |\bm{p}_1-\bm{p}_2|>k_F,\\& |\bm{k}+\bm{p}_2|>k_F
\raisetag{12pt}
\label{secondcase}
\end{split}
\end{align}
We make the following change of variables:
$$ \bm{p}'_1 = \bm{p}_1-\bm{p}_2,\, \bm{p}'_2 = \bm{k}+\bm{p}_2
\Rightarrow \bm{p}_1 = \bm{p}'_1+\bm{p}'_2 - \bm{k} , \, \bm{p}_2 =- \bm{k}+\bm{p}'_2$$
$\bm{1}.$ First we treat the case in \ref{firstcase} where $\bm{p}'_1$ and $\bm{p}'_2$ are both bounded and therefore we get the following inequalities:
$$|\bm{k}|>k_F, \, | \bm{p}'_1+\bm{p}'_2 - \bm{k}|>k_F, \, |\bm{p}'_1|<k_F,\, |\bm{p}'_2|<k_F$$
We observe from the second inequality that since $|\bm{k}|$ can go to infinity and $|\bm{p}'_1|$ and $|\bm{p}'_2|$ are bounded then the latter two are negligible for $|\bm{q}|\rightarrow \infty$, so we set $\bm{p}'_1=\bm{p}'_2=0$ in the following and we replace $\omega = -q^2/2m_i$, we notice that the Heaviside function $\theta(|\bm{k}+\bm{q}|>k_F)=1$ for all values of $\bm{k}$ here:
\begin{align}
	\begin{split}
&\sum_{k, \bm{p}_1, \bm{p}_2}
\frac{1}{(\omega_1-k^2/2m+i0^+)^2(\omega_1+k^2/2m-i0^+)}
\\&
\frac{1}{(\omega_1-q^2/2m_i-(\bm{k}+\bm{q})^2/2m+i0^+)}
+\frac{1}{2\epsilon_{p_2}}
	\end{split}
\end{align}
The function has 3 poles with respect to $\omega_1$ and we integrate over the upper half side of the complex plane:
\begin{align}
	\begin{split}
&\sum_{\bm{k}, \bm{p}_1, \bm{p}_2}
\frac{1}{
(-q^2/2m_i-k^2/2m-(\bm{k}+\bm{q})^2/2m+i0^+)}
\\&
\frac{1}{(-k^2/m+i0^+)^2}
+\frac{1}{2\epsilon_{p_2}}
	\end{split}
\end{align}
The two integrals over $\bm{p}_1$ and $\bm{p}_2$ give each a factor equal to the density of the Fermi gas $n = k_F^3/6\pi^2$. Then we perform a variable change $\bm{k} = q \bm{p} $:
\begin{align}
	\begin{split}
&\frac{1}{q^3} \int_0^{\infty}\frac{\text{d}\bm{k}}{(2\pi)^3}
\frac{-i}{(-k^2/m)^2
(q^2/2m_i+k^2/2m+(\bm{k}+\bm{q})^2/2m)}
	\end{split}
\end{align}
$\bm{2}.$ Second, we treat the case in \ref{secondcase} where $\bm{p}'_1$ and $\bm{p}'_2$ are both bounded and therefore we get the following inequalities:
$$|\bm{k}|<k_F, \, |\bm{p}_1|<k_F, \, |\bm{p}_1-\bm{p}_2|>k_F, \, |\bm{k}+\bm{p}_2|>k_F$$
We see that $\bm{k}$ and $\bm{p}_1$ are bounded so they go to $0$ and we get:
\begin{align}
	\begin{split}
&\sum_{k, \bm{p}_1, \bm{p}_2}
\frac{\theta(|\bm{k}+\bm{q}|>k_F)}{(\omega_1-E_{\bm{k}}+i\eta_1)^2(\omega_1+E_{\bm{p}_1}-E_{\bm{p}_1-\bm{p}_2}-E_{\bm{k}+\bm{p}_2}+i\eta)}
\\&
\frac{1}{(\omega_1+\omega-E_{\bm{k}+\bm{q}}+i0^+)}
+\frac{1}{2\epsilon_{p_2}}
\label{eq:1.6}
	\end{split}
\end{align}
The function has 4 poles with respect to $\omega_1$ and we integrate over the upper half side of the complex plane:
\begin{align}
	\begin{split}
&i\frac{d}{d\omega_1}
\frac{1}{
(\omega_1-q^2/2m_i+i0^+)(\omega_1-p_2^2/2m_i+i0^+)}
|_{\omega_1 = i0^+}\\&
=\frac{-i}{(q^2/2m_i)^2(p_2^2/m)}+
\frac{-i}{(q^2/2m_i)(p_2^2/m)^2}
\frac{1}{(-k^2/m+i0^+)^2}
	\end{split}
\end{align}
By integrating this expression and in addition to the result of the first case we get for this diagram:
$$ig_0^2 (\frac{k_F}{6\pi^2})^2 m \kappa_I(\eta) \frac{1}{q^3}$$
with $\eta = m_i/m$. This is the first contribution that appears in Eq. \ref{Eq:AsymptoticChi}.

\begin{figure}[h]
\centering
\includegraphics[width=0.36\textwidth]{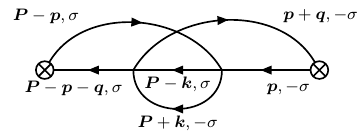}
\caption{Another contribution to density-density response function to second order in the fermion-fermion interaction that shows a $1/q^3$ divergent term.
}
\label{pert2}
\end{figure}

\begin{figure}[h]
\centering
\includegraphics[width=0.36\textwidth]{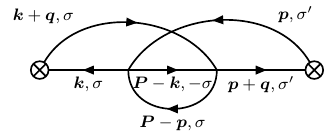}
\caption{Another contribution to density-density response function to second order in the fermion-fermion interaction that shows a $1/q^3$ divergent term. 
}
\label{pert3}
\end{figure}

Other diagrams where the two interactions happen between the fermions resulting from the interaction with the impurity. These diagrams do not contribute except for the two diagrams shown in Figs. \ref{pert2} and \ref{pert3}, the first one is the ladder diagram to second order while the other is the crossed ladder diagram to second order. 
The diagram in Fig. \ref{pert2} gives the contribution in $\kappa_{II}(\eta)$. The total momentum of the interaction vertices diverges so we have to plug in the full interaction vertex and this leads to calculating the diagram in Fig. \ref{gen2} which we do in the following. Thus, the diagram has the following contribution 
\begin{align}
	\begin{split}
&\frac{2i}{V} 
\sum_{P,p} 
\frac{\theta(|\bm p|> k_F)}
{\omega_1 - E_{\bm p}+i\eta}
\frac{\theta(|\bm P - \bm p-\bm q|> k_F)}
{\Omega-\omega_1 - \omega - E_{\bm P -\bm p - \bm q}+i\eta}
\\
&\frac{\theta(|\bm p+\bm q|> k_F)}
{\omega + \omega_1 - E_{\bm p+ \bm q}+i\eta}
\frac{\theta(|\bm P - \bm p|> k_F)}
{\Omega- \omega_1 - E_{\bm P -\bm p}+i\eta} 
\Gamma (P)
\label{eq:FULLEXPRESSIONCHI11R}
\raisetag{30pt}
\end{split}
\end{align}
where $P=(\bm P, \Omega), p = (\bm p, \omega_1)$ are internal four momenta and $ q = (\bm q, \omega)$ is the four momentum of the impurity.\\
We can write an expression for the Bethe-Salpeter equation for $\Gamma$, recalling that at $T=0$, Feynman rules add a factor $i$ in front of the recursive part, as follows:
\begin{align}
	\begin{split}
		&\Gamma^{-1} (\bm P, \Omega) = g_0^{-1}
		-\sum_{|\bm p_1|>k_F}
		\frac{\theta(|\bm P - \bm{p}_1|> k_F)}{\Omega-E_{\bm p_1} - E_{\bm P -\bm p_1}+i\eta}
		\\&+\sum_{|\bm p_1|<k_F}
		\frac{\theta(|\bm P - \bm{p}_1|< k_F)}{\Omega-E_{\bm p_1} - E_{\bm P -\bm p_1}-i\eta}
		\label{eq:vertexgamma}
	\end{split}
\end{align}
with $g_0^{-1} = g^{-1} -\sum_{p_1}\frac{2m^* }{p_1^2}$.
We need to evaluate the expression at the frequency value $\omega=-\epsilon_{\bm q}$ which corresponds to the impurity's kinetic energy. Then we take the $|\bm q| \rightarrow \infty$ limit. For that we can write Eq. \eqref{eq:FULLEXPRESSIONCHI11R} as:
\begin{align}
	\begin{split}
	\int\frac{d^3\bm P}{(2\pi)^3} \int_{-\infty}^{\infty} \frac{d\Omega}{2\pi} \Gamma (\bm P, \Omega) \, \, F(\Omega, \bm P,q,k_F,m,m_i)
	\label{eq:FULLEXPRESSIONCHI11R3}
\raisetag{0pt}
\end{split}
\end{align}
where the function $F$ is given by:
\begin{align}
	\begin{split}
 F &= \int \frac{d^3 \bm p}{(2 \pi)^3} 
 \frac{\theta(|\bm p+\bm q|> k_F)\theta(|\bm P - \bm p|> k_F)}{(\Omega -\epsilon_{\bm q} - E_{\bm P -\bm p} - E_{\bm p+ \bm q}+i\eta)}
\\& \frac{\theta(|\bm p|> k_F)\theta(|\bm P - \bm p-\bm q|> k_F)}
{(\Omega - E_{\bm P -\bm p} - E_{\bm p} + i\eta)(\Omega-E_{\bm p+ \bm q} - E_{\bm P -\bm p - \bm q}+i\eta)}
	\label{eq:FUNCTIONF}
\raisetag{50pt}
\end{split}
\end{align}
We note that $F$ is holomorphic in the upper half of the complex plane with respect to $\Omega$. We split the complex function $ \Gamma (P)$ into a sum of an advanced and a retarded function: $\Gamma (P) = \Gamma^R (P) + \Gamma^A (P)$. 
\\The function $\Gamma^R (P)$ is holomorphic in the upper-half of the complex plane and  $\Gamma^A (P)$ is holomorphic in the lower half of the complex plane. With that we find that only $\Gamma^A (P)$ will contribute in Eq. \eqref{eq:FULLEXPRESSIONCHI11R3} in order for the integrand to not be holomorphic in the lower-half plane and integrate to zero with respect to $\Omega$. Now, we rescale $|\bm p|$ by $|\bm q|$:
$$ \frac{(2m)^3}{q^3} F(\frac{\Omega}{q^2/(2m)},\frac{|\bm P|}{|\bm q|},1, \frac{k_F}{q},1,\frac{m}{m_i})$$
By studying the behavior of the function $\Gamma (\bm P, \Omega)$ at $|\bm P|, \Omega \rightarrow \infty$ we find that its limit is zero. We prove that by noticing that the second sum in Eq. \eqref{eq:vertexgamma} is zero when $|\bm P|\rightarrow \infty$, we write the Heaviside function in the first sum as $1-\theta(|\bm P - \bm{p}_1|< k_F)$ and the second term subsequently goes to zero for $|\bm P|\rightarrow \infty$. We can calculate the remaining sum
 to find that $\Gamma^{-1} (\bm P, \Omega)$ diverges for $|\bm P|, \Omega \rightarrow \infty$.
 \\ As a result, in $F$ we can replace the first two arguments in the last expression by zero at lowest order and we find the diagram to be 
$ \underset{|\bm q|\rightarrow \infty}{\propto}  \frac{A}{q^3}$
with A given by:
$$A = \int\frac{d^3\bm P}{(2\pi)^3} \int_{-\infty}^{\infty} \frac{d\Omega}{2\pi} \gamma^A_{\uparrow,\downarrow} (\bm P, \Omega) \, \, (2m)^3 F(0, 0,1,0,m,m_i)
$$
By definition we have $$  \int_{-\infty}^{\infty} \frac{d\Omega}{2\pi} \Gamma^A (\bm P, \Omega)=\Gamma (\bm P, t=0^-)$$
and equivalently:
$$  \int\frac{d^3\bm P}{(2\pi)^3} \Gamma (\bm P, t=0^-)=\Gamma (\bm r = \bm 0, t=0^-)$$
The last expression can be related to the two-body contact as shown in \cite{rossi2018,houcke2019} (the factor $i$ comes from the zero temperature formalism):
$$\mathcal{C}_2 = i m^2\,\Gamma (\bm r = \bm 0, t=0^-)$$
In this context the contact $\mathcal{C}_2$ will help us identify diverging terms as it appears as a prefactor for these terms. \\
With that the coefficient of the divergent term becomes:$$A = - 8 \, m \, \mathcal{C}_2 F(0, 0,1,0,m,m_i)$$
This gives one of the contributions to the divergent term in Eq. \eqref{eq:EnerPolMat}. With the notations used in \cite{Pierce19few} we calculate the function $F(0, 0,1,0,m,m_i)$ and we find:
$$F(0, 0,1,0,m,m_i) = \frac{m^3}{\pi^2} \kappa_{II}(\eta)$$
where $\eta = m_i/m$ and $\kappa_{II}(\eta) =-\frac{\pi}{2} \eta \arctan\left( \frac{1}{\sqrt{\eta(\eta+2)}}\right)$. 

The final diagram contributes to the $\kappa_{III}(\eta)$ term.
For the diagram in Fig. \ref{pert3}, the calculation takes the same steps but we have to pay attention that the Fermi-Fermi vertex on the right side of the diagram cannot be summed perturbatively but should be replaced by the dressed vertex $\Gamma$ as in the previous diagram. 
The relevant momentum of this vertex diverges and therefore we replace $\Gamma(P)$ by $-4\pi/(m\sqrt{-m(\Omega-{\bf P}^2/(4 m)+i\,0^+)})$. We write the following expression, dominant in the $q\to\infty$ limit
\begin{align}
		\begin{split}
&\sum_{P,k,p,\sigma,\sigma'} 
G_{0,\sigma}(k)
G_{0,\sigma}(k+q)
G_{0,-\sigma}(P-k)
\\&
G_{0,-\sigma'}(P-p)
G_{0,\sigma'}(p)
G_{0,\sigma'}(p+q)
\Gamma(P+q)
\Gamma(P)
\\&\simeq -4\sum_{P,{\bm k},{\bm p}} 
\frac{4\pi \,\Gamma(P) \theta(|\bm{k}|>k_F)}{m\sqrt{-m(\Omega - q^2/2m_i-E_{\bm{k}+\bm{q}} +i0^+)}}
\\&\frac{\theta(|\bm{k}+\bm{q}|>k_F)\theta(|\bm{P}-\bm{k}|>k_F)\theta(|\bm{P}-\bm{p}|>k_F)}{\Omega-E_{\bm{P}-\bm{k}}-q^2/2m_i-E_{\bm{k}+\bm{q}}+i0^+}
\\&
\frac{\theta(|\bm{p}|>k_F)\theta(|\bm{p}+\bm{q}|>k_F)}{\Omega - E_{\bm{P}-\bm{k}}-E_{\bm{k}}+i0^+}
\raisetag{20pt}
	\end{split}
\end{align}
Following the same steps as the other two calculations detailed above we get the following result:
$$\frac{128\, \pi\, m\, \mathcal{C}_{2}}{n}\sqrt{\frac{\eta}{\eta+2}}\left(J_{c}(\eta)\right)^{2}
\frac{1}{q^3}$$

\section{Two interaction vertices subdominant diagram}
\label{appendixB}
A second diagram with two bold two-particles vertices $\Gamma$ is shown in Fig. \ref{apdx1}. The analytic expression is (global sign is irrelevant, since we argue that it gives a subdominant contribution in the $q\to \infty$ limit): 
\begin{figure}[h]
\centering
\includegraphics[width=0.48\textwidth]{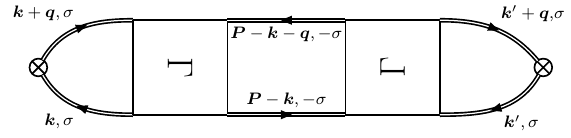}
\caption{The diagram with two bold interaction vertices and bold propagators for the fermions.
}
\label{apdx1}
\end{figure} 
\begin{align}
	\begin{split}
    \chi_{1\,d}(\mathbf{q}, \varepsilon^{(i)}_q)
    &=\frac{\pm i}{n}\sum_{\sigma,\sigma'}\int_{0}^{+\infty}dt\int_{\mathbb{R}^{4}} \prod_{i=1}^{4} d t_{i} \int \frac{d^{3} k d^{3} k^{\prime} d^{3} P}{(2 \pi)^{9}}\\
    &e^{-i\frac{q^2}{2\,m_i}t}
    G_{\sigma}\left(\mathbf{k}-\mathbf{q}, t_{1}\right)
     G_{\sigma}\left(\mathbf{k}, -t_{2}\right)\\
     &
     \Gamma\left(\mathbf{P}, t_{2}-t_{1}\right)
     G_{-\sigma}\left(\mathbf{P}-\mathbf{k}+\mathbf{q}, t_{1}-t_{4}\right)  \\&
  G_{-\sigma}\left(\mathbf{P}-\mathbf{k}, t_{3}-t_{2}\right)
   \Gamma\left(\mathbf{P}+\mathbf{k}'-\mathbf{k}, t_{4}-t_{3}\right)
   \\&
  G_{\sigma}\left(\mathbf{k}', t_{3}-t\right)  
   G_{\sigma}\left(\mathbf{k}'-\mathbf{q}, t-t_{4}\right)
  \label{eq:chi1c}
  \raisetag{10pt}
\end{split}
\end{align}
If we assume that all $G$'s are retarded with momenta which are large {\it i.e.} of the order $q$, this gives seven conditions for the different times: $t_1>0$, $t_2<0$, $t_3>t_2$, $t_1>t_4$, $t>t_4$, $t_3>t$ and $t>0$.
The first, sixth and seventh conditions imply the third condition, so we have six conditions. 
If we assume that $|\mathbf{P}|>\Lambda$ is large, this means that the dominant contribution is for $t_2>t_1$, which is inconsistent with the conditions. 
Therefore we assume $|\mathbf{P}|<\Lambda$. We find that $|\mathbf{P}+\mathbf{k}'-\mathbf{k}|\gg\Lambda$, and the dominant contribution is for $t_4>t_3$, which is inconsistent with previous conditions. As a conclusion, we find that this diagram is not dominant, compared to the diagram for $\chi_{1\,c}$.

We can recover this result in another manner. We assume that all the fermionic wavevectors of $G$'s are of order $q$ and are large compared to $k_{typ}$. This means that at lowest order all the $G$'s are retarded. Consider the time loop : $0\rightarrow t_1\rightarrow t_2 \rightarrow 0$. Due to the retarded nature of the two $G$'s in this loop, the time difference in $\Gamma$ in this loop must be negative. This means that at lowest order, the wavevector of the $\Gamma$ in the loop cannot be large. We come to the same conclusion for the wave vector of the second $\Gamma$, by considering the time loop $t\rightarrow t_4\rightarrow t_3 \rightarrow t$. Following the same procedure as before, we see that there is only one independent wavevector which is large. The integral on this large wavevector gives a factor $q^3$, while the integrals on the five time differences give a factor $(q^{-2})^5=q^{-10}$. This gives finally a subdominant  contribution of order $q^{-7}$.

\section{Subdominant diagrams with more than three vertices}
\label{appendixC}
In this section, we give arguments which justify that bold diagrams for the density-density response function with three or more two-particles vertices give a subdominant contribution in the $q\to\infty$ limit.
We assume that dominant
contributions in integrals come from high wavectors ({\it i.e.} larger than $k_{typ}$)
of Green's functions. The Green's functions are then 
replaced by free particle Green's functions (indeed for negative time differences, due to Property 3, the Green's function tends to zero like the wavevector to the power $-4$).
Consider a diagram with $M$ two-particles vertices $\Gamma$. For $M=0$ the diagram is shown in Fig. \ref{gen1}, for $M=1$ it is shown in Fig. \ref{gen2}. The two diagrams for $M=2$ are shown in Figs. \ref{gen3} and \ref{apdx1}.
The $q$ dependence comes from three contributions. 
The first contribution is a "phase space" contribution: one integrates on internal wavevectors which tend to infinity and scale like $q$. Each three dimensional integration gives a factor $q^3$.
We denote $N_1$ the number of such independent 
wavevectors. The integrations give a factor $q^{3\,N_1}$.
The second contribution comes from integration on "small" positive
({\it i.e} smaller than $t_{typ}$) time differences entering
Green's functions. According to our hypothesis, the 
wavevectors are of order $q$ and each time integration gives a factor of order $q^{-2}$. We denote $N_2$ the number of independent time differences entering Green's functions. Theses integrations give an factor $q^{-2 N_2}$.
The third contribution comes from integration on small positive time 
differences of $\Gamma$'s, if the wavevector is "large". The time integration 
gives a factor $q^{-1}$ (see section \ref{sec2vert}). We denote $N_3$ the 
number of such time differences and wavevectors. These time integrations 
give a factor $q^{-N_3}$. In total, the contribution of a diagram with numbers
$N_1$, $N_2$ and $N_3$ scales like $q^{\alpha}$, with
\begin{equation}
\alpha=3\,N_1-2\,N_2-N_3
 \label{eq:eqalpha}   
\end{equation}
As an example, for the diagram in Fig.\ref{gen1}, we have $M=0$, $N_1=0$, $N_2=1$ and $N_3=0$, and $\alpha=-2$. For the diagram in Fig.\ref{gen2}, we have $M=1$, $N_1=1$, $N_2=3$ and $N_3=0$, and $\alpha=-3$. For the diagram in Fig.\ref{gen3}, we have $M=2$, $N_1=2$, $N_2=4$ and $N_3=1$, and $\alpha=-3$.
The diagram in Fig. \ref{apdx1} is subdominant. Indeed, we have seen in section \ref{appendixB} that $N_1=1$, $N_2=5$, $N_3=1$ and $\alpha=-7$.

We now consider the general case, with $M\geq 3$ vertices $\Gamma$. There are $M+1$ independent wave vectors.
$N_3$ wavevectors of $\Gamma$’s are high and therefore there are $N’_3=M-N_3$ low wave vectors for $\Gamma$’s. The total number of independent wavevectors which are high is therefore
$N_1=M+1-N’_3=N_3+1$. There are $2\,M+1$ independent time differences. Among these time differences, $N_3$ are assigned to $\Gamma$’s with high momenta. We assume that all the remaining ones are assigned to $G$’s which have high momenta and are retarded. This gives $N_2=2\,M+1-N_3$. We find
\begin{equation}
\alpha=4\,N_3-4\,M+1.
\label{eq:alphaN3}
\end{equation}
Using this formula, we recover the results we obtained for $M=1$ and $M=2$. Indeed, in these cases, $N_3=M-1$ and $\alpha=-3$. If $N_3\leq M-2$, we find $\alpha\leq -7$.

For $M\geq 3$, we argue that $N_3\leq M-2$, or equivalently that $N'_3$, the number of advanced $\Gamma$'s,  is larger than $2$. Indeed, for $M=3$, by inspection of all the possible bold diagrams, we found that at least $2$ $\Gamma$'s are advanced. This is due to time loops that involve one or two $\Gamma$'s and Green's functions that are retarded, and we expect this will occur in general.
 $N_3\leq M-2$ means, using Eq. (\ref{eq:alphaN3}), that $\alpha\leq -7$, and we conclude that these diagrams give subdominant contributions.
